\documentclass[twoside]{article}

\usepackage[accepted]{aistats2026}
\usepackage{enumitem}
\usepackage{amsthm} 
\usepackage[hyphens]{url}  
\usepackage{graphicx} 
\usepackage{natbib}  
\usepackage{xcolor}
\usepackage{caption} 
\usepackage{dsfont}
\usepackage{booktabs} 
\usepackage{multirow} 
\usepackage{algorithm}
\usepackage{amssymb}
\usepackage{natbib}
\usepackage{hyperref}
\usepackage{algpseudocode}
\newtheorem{theorem}{Theorem}
\newtheorem{lemma}{Lemma}

\newtheorem{definition}{Definition}

\newtheorem{remark}{Remark}

\begin{document}

%

%
\runningauthor{Alkhouri, Wu, Yu, Liu, Wang, Velasquez}

\twocolumn[

\aistatstitle{A Scalable Lift-and-Project Differentiable Approach For the Maximum Cut Problem}

\aistatsauthor{%
  \begin{tabular}{c}
    Ismail Alkhouri$^{*1,2}$, 
    Mian Wu$^{*3}$, 
    Cunxi Yu$^{4}$, 
    Jia Liu$^{3}$, 
    Rongrong Wang$^{\dagger5,6}$,
    Alvaro Velasquez$^{\dagger7}$
  \end{tabular}
}

\aistatsaddress{%
  $^{1}$X-Computational Physics - Los Alamos National Laboratory \\
  $^{2}$Michigan Institute for Computational Discovery \& Engineering - University of Michigan \\
  $^{3}$Department of Electrical \& Computer Engineering - Ohio State University \\
  $^{4}$Department of Electrical \& Computer Engineering - University of Maryland \\
  $^{5}$Department of Computational Mathematics, Science, \& Engineering - Michigan State University \\
  $^{6}$Department of Mathematical Sciences - Michigan State University \\
  $^{7}$Department of Computer Science - University of Colorado, Boulder 
  }
 ]

\begin{abstract}

We propose a scalable framework for solving the Maximum Cut (MaxCut) problem in large graphs using projected gradient ascent on quadratic objectives. Our approach is differentiable and leverages GPUs for gradient-based optimization. It is not a machine learning method and does not require training data. Starting from a continuous relaxation of the classical quadratic binary formulation, we present a parallelized strategy that explores multiple initialization vectors in batch. We analyze the relaxed objective, showing it is convex and has fixed-points corresponding to local optima—particularly at boundary points—highlighting a key challenge in non-convex optimization. \textcolor{black}{To improve exploration}, we introduce a lifted quadratic formulation that over-parameterizes the solution space. We also provide a theoretical characterization of these lifted fixed-points. Finally, we propose DECO, a dimension-alternating algorithm that switches between the unlifted and lifted formulations, combined with importance-based degree initialization and a population-based evolutionary hyper-parameter search. Experiments on diverse graph families show that our methods attain comparable or superior performance relative to recent \textcolor{black}{neural networks} and GPU-accelerated sampling approaches. 
\end{abstract}


\section{INTRODUCTION}\label{sec: intro}


A fundamental connection between Combinatorial Optimization Problems (COPs) and NP-hardness was established in the influential work of Karp in~\citep{karp1972reducibility}, highlighting their intrinsic computational difficulty. The paper also introduced the concept of reducibility among NP-complete COPs, allowing their relative complexity to be examined through reductions between problems.


%

Although there could be a straightforward reduction between some COPs -- such as reducing the Maximum Cut (MaxCut) and the Not-all-equal 3-satisfiability (NAE 3SAT) \citep{moret1988planar} --  which allows a solution for one problem to be used to solve another, other COPs differ significantly. For example, there does not exist a direct reduction between MaxCut and the Kidney Exchange Problem (KEP) \citep{mcelfresh2019scalable}.

In this work, we focus on the MaxCut problem, a fundamental COP with wide-ranging applications. These include, but are not limited to, interference management in wireless networks~\citep{gu2024graph}, physical design in VLSI circuits~\citep{liers2011via}, and ensuring consistency in phylogenetic tree construction in biology~\citep{4015375}. 



The MaxCut problem partitions the set of nodes $V$ of graph $G=(V,E)$ into two disjoint sets, such that the number of edges crossing between the two sets is maximized. Several methods have been proposed to tackle the MaxCut problem, including the ones based on Integer Linear Programming (ILP)~\citep{MaxCut_bnb_lu2021} and Quadratic Unconstrained Binary Optimization (QUBO) \citep{glover2019tutorial}. However, as ILP and QUBO formulations rely on integer variables, they do not scale well (\textcolor{black}{in the worst case}), often incurring high computational costs on large and/or dense graphs. Heuristic approaches, such as Breakout Local Search (BLS) \citep{benlic2013breakout}, offer faster solutions but generally lack theoretical guarantees and often require problem-specific and graph-specific tuning. A widely known approximation algorithm based on semidefinite programming (SDP) was introduced by Goemans and Williamson \citep{goemans1995improved}, providing strong theoretical approximation guarantees. However, the SDP solution must be rounded to obtain a valid cut, and such rounding may result in suboptimal solutions.

Recent years have seen a surge in learning-based approaches for MaxCut and other COPs, including supervised, unsupervised, and reinforcement learning methods~\citep{bother2022s, pmlr-v119-ahn20a, opt_gnn}. While these methods have shown promise (in terms of obtaining fast solutions at inference), they typically require a large amount of labeled or unlabeled training graphs. As a result, \textit{many} of them often generalize poorly to out-of-distribution (OOD) instances and struggle to scale to larger graphs, even from the same distribution (see the results of DIFUSCO \citep{sun2023difusco} on the large graphs in \citep{pCQO}). To address these limitations, training-data-free (or data-less) graph neural network (GNN) methods have been proposed~\citep{schuetz2022combinatorial, pignn_cra}, where the GNN parameters are optimized using a relaxed QUBO formulation over the single graph that defines the given problem instance. However, some of these methods remain constrained by input encodings that hinder scalability and typically only perform well on relatively sparse graphs.

An alternative class of data-less approaches—those based on GPU-accelerated optimization or sampling methods—has recently emerged~\citep{pCQO, sun2023revisiting, resco, velasquez2025dataless}. These methods avoid both the need for training data and the graph encoding bottlenecks of GNN-based techniques, and have demonstrated competitive performance, even against state-of-the-art (SOTA) problem-specific heuristics. However, they often rely on hand-tuned hyper-parameters (typically a small set of scalar values) to achieve strong performance.

In parallel, the increasing capabilities of modern GPUs, as well as advances in distributed and parallel computing, have enabled solvers based on differentiable frameworks—such as the method based on Gumbel-softmax presented in~\citep{pmlr-v235-liu24al}—to outperform traditional CPU-bound solvers, including commercial packages like Gurobi \citep{Gurobi} and CP-SAT \citep{cpsatlp}, on large-scale scheduling COPs.

Motivated by the benefits of GPU-based parallel computation and the flexibility of training-data-free optimization \citep{velasquez2025dataless}, this paper proposes two data-less differentiable quadratic approaches for the MaxCut problem, along with an alternating optimization algorithm. The first formulation is a relaxed QUBO, while the second introduces a lifted quadratic optimization. The contributions of the paper are summarized as follows:

 \begin{enumerate}

    \item \textbf{Relaxed and Lifted Formulations}: We propose a relaxed unlifted QUBO formulation and introduce a lifted quadratic formulation \textcolor{black}{for improving exploration}.
    
    \item \textbf{Theoretical Insights}: We provide theoretical insights into the relaxed unlifted objective, including convexity of the objective \textcolor{black}{and} the graph Laplacian with its null space. We identify two types of fixed-points including boundary points that explain why local minima arise in those settings. Furthermore, we theoretically characterize the MaxCut lifted fixed-points.

    \item \textbf{Scalable Parallelized Algorithms}: Building on these formulations and \textcolor{black}{the} fixed-point characterizations, we develop three GPU-parallelized scalable algorithms, including a dimension-alternating scheme that switches between the unlifted and lifted formulations. 

    \item \textbf{Initialization and Hyper-Parameter Search}: To mitigate sensitivity to initial vectors and reduce manual \textcolor{black}{hyper-parameter} tuning, we adopt an importance-based initialization strategy and employ a population-based evolutionary algorithm for parameter search.
    

      \item \textbf{Extensive Empirical Evaluation}: On large (\textcolor{black}{synthetic, real-world, and well-known}) graph datasets, we show that our \textcolor{black}{algorithms}—combined with batching, degree-based initialization, and hyper-parameter search—achieve competitive MaxCut solutions at scale. Our results suggest that differentiable, parallelized, and training-data-free approaches can rival or surpass data-intensive and recent sampling-based methods.

 \end{enumerate}

\section{PRELIMINARIES}\label{sec: prel}

\paragraph{Notations:} Let $G=(V,E)$ be an undirected graph, where $V$ is the set of nodes and $E \subset V \times V$ is the set of edges. The number of nodes and edges are denoted by $|V| = n$ and $|E| = m$, respectively, where $|\cdot|$ indicates the cardinality of a set. For a node $v\in V$, its degree $\textrm{d}_v$ is the number of edges connected to $v$, and the maximum degree in the graph is denoted by $\Delta$. The diagonal degree matrix is $\mathbf{D}$, where $\mathbf{D}_{v,v} = \textrm{d}_v$. The symmetric adjacency matrix of the graph is denoted by $\mathbf{A} \in \{0,1\}^{n \times n}$, where $\mathbf{A}_{u,v} = 1$ if $(u,v)\in E$ and $0$ otherwise. The identity matrix is denoted by $\mathbf{I}$. The trace of a matrix $\mathbf{A}$ is denoted by $\mathrm{tr}(\mathbf{A})$. For any positive integer $n$, we use $[n] := \{1, \ldots, n\}$. The vector of all ones of size $n$ is denoted by $\mathbf{e}_n$ and the vector of all zeros is denoted by $\mathbf{0}_n$. We also use $\mathds{1}\{\cdot\}$ to denote the indicator function, which returns $1$ if its argument is True and $0$ otherwise. 

We now formally define the NP-hard problem of finding a maximum cut in a graph.

\begin{definition}[Maximum Cut (MaxCut)] 
Given an undirected graph $G=(V,E)$, the goal of the MaxCut problem is to partition the nodes into two disjoint sets $S$ and $\bar{S} = V \setminus S$ such that the number of edges crossing the cut (i.e., with one endpoint in $S$ and the other in $\bar{S}$) is maximized.
\end{definition}

Given any set $S \subset V$, the cut value is defined as:
\begin{align}\label{eqn: cut value from set}
    \textrm{Cut}(S) = \sum_{v\in S} \sum_{u\in V \setminus S} \mathds{1}\{(v,u)\in E\}\:.
\end{align}
Let $\mathbf{z} \in \{0,1\}^n$ be a binary vector where each entry $\mathbf{z}_v$ corresponds to node $v \in V$, and let  $\mathbf{y}_{v,u}$ corresponds to an edge $(v,u) \in E$ stacked in an $m$-dimensional binary vector. Then, the MaxCut problem can be formulated as the following ILP:
\begin{align}\label{eqn: cut ILP}
    \max_{\mathbf{z}\in\{0,1\}^{n},~~\mathbf{y}\in\{0,1\}^{m}} \quad & \sum_{(v,u)\in E} \mathbf{y}_{v,u} \\
    \text{s.t.} \quad & \mathbf{y}_{v,u} \leq \mathbf{z}_{v} + \mathbf{z}_{u}, \nonumber \\
    & \mathbf{y}_{v,u} \leq 2 - \mathbf{z}_{v} - \mathbf{z}_{u}, \quad \forall (v,u) \in E \:. \notag
\end{align}

Let $\mathbf{y}^*$ be the optimal solution to \eqref{eqn: cut ILP}. Then, the optimal cut value is $\sum_{(v,u)\in E} \mathbf{y}_{v,u}^*$. This ILP involves $n + m$ binary variables. An alternative formulation that uses only $n$ binary variables is the following QUBO, where the optimal cut value is equal to the value of the objective:
\begin{align}\label{eqn: cut qubo}
    \max_{\mathbf{z} \in \{-1,1\}^{n}} \frac{1}{2} \sum_{(v,u)\in E} (\mathbf{z}_v - \mathbf{z}_u)^2\:.
\end{align}
Since both \eqref{eqn: cut ILP} and \eqref{eqn: cut qubo} are binary programs, they scale poorly with increasing $n$ and $m$. To address this, lifting-based relaxations have been explored—most notably, the following semidefinite programming (SDP) relaxation of MaxCut:
\begin{align}\label{eqn: cut SDP}
    \max_{\mathbf{W} \in \mathbb{R}^{n \times n}} \quad & \frac{1}{2} \sum_{(v,u)\in E} (1 - \mathbf{W}_{v,:}^{\top} \mathbf{W}_{u,:}) \nonumber \\
    \text{s.t.} \quad & \mathbf{W}_{v,v} = 1, \quad \forall v \in V, \\
    & ~~~~\mathbf{W} \succeq 0\:, \notag
\end{align}
where $\mathbf{W}_{v,:}$ denotes the $v$-th column of $\mathbf{W}$. While \eqref{eqn: cut SDP} is convex and solvable in polynomial time, it has two primary limitations: (i) a rounding procedure is required to map the solution $\mathbf{W}^*$ to a feasible cut in the original graph, often yielding suboptimal results; and (ii) scalability issues arise as $n$ increases, due to the positive semidefinite constraint $\mathbf{W} \succeq 0$. To address the first limitation, Goemans and Williamson \citep{goemans1995improved} proposed a rounding algorithm—known as the Goemans-Williamson (GW) algorithm—that samples a random hyperplane through the origin and partitions the node embeddings accordingly. This method guarantees, in expectation, a cut value that is at least $0.878$ times the optimal value.



\section{PROPOSED OBJECTIVE FUNCTIONS}\label{sec: proposed functions}

In this section, we introduce a relaxed quadratic objective for MaxCut, describe its optimization using gradient ascent, characterize fixed-points, propose a lifted quadratic formulation to improve exploration, and characterize the lifted fixed-points. 

\subsection{Differentiable Quadratic Optimization}\label{sec: proposed functions relaxed qubo}

Let $\mathbf{x}\in [-1,1]^n$ be a vector where each entry $\mathbf{x}_v$ corresponds to node $v\in V$. A relaxed version of the QUBO formulation in \eqref{eqn: cut qubo} can be written as:
\begin{align}\label{eqn: cut quco unlifted}
    \max_{\mathbf{x}\in [-1,1]^{n}} \quad f(\mathbf{x}) 
    & := \frac{1}{2}\sum_{(v,u)\in E}(\mathbf{x}_{v}-\mathbf{x}_{u})^2 \nonumber \\
    & ~= \mathbf{x}^{\top}(\mathbf{D}-\mathbf{A})\mathbf{x} 
    =  \mathbf{x}^{\top}\mathbf{L}\mathbf{x} \:, \tag{\texttt{QUCO}}
\end{align}
where $\mathbf{L}$ is the graph Laplacian. An equivalent form is derived by expanding the objective:  
\begin{align*}
    \frac{1}{2}\sum_{(v,u)\in E}(\mathbf{x}_{v}-\mathbf{x}_{u})^2 = \sum_{v\in V}\mathrm{d}_v \mathbf{x}^2_v - \sum_{(u,v)\in E} \mathbf{x}_v \mathbf{x}_u \:.   
\end{align*}
We term this formulation as QUadratic box-Constrained Optimization \eqref{eqn: cut quco unlifted}. Notably, if $\mathbf{x}$ lies at the boundary (i.e., entries in $\{-1,1\}^n$), then the objective function is equivalent to the objective of the binary program in \eqref{eqn: cut qubo}. 

Let $\mathbf{z}\in \{0,1\}^n$ represent the binarized version of $\mathbf{x}\in [-1,1]^n$, i.e., 
\begin{equation}\label{eqn: cut from binary vec}
\mathbf{z}_v = \mathds{1}\{\mathbf{x}_v>0\},~\text{then the cut value is }~~\mathbf{z}^{\top}\mathbf{L}\mathbf{z}\:.    
\end{equation}
The recent study in \citep{pCQO} showed that using projected gradient descent with a relaxed box-constrained version of the Maximum Independent Set (MIS) QUBO (on an augmented formulation) yields better and/or competitive solutions against several SOTA learning-based and sampling-based methods. Motivated by this, in this paper, we first propose to solve \eqref{eqn: cut quco unlifted} using projected gradient ascent. 


Let the gradient of $f$ be denoted as 
\begin{equation}\label{eqn: grad of unlifted}
  \mathbf{g}(\mathbf{x}) := \nabla_{\mathbf{x}}f(\mathbf{x}) = \mathbf{L}\mathbf{x}.  
\end{equation}
%
Starting from an initial point $\mathbf{x}$, the updates in \eqref{eqn: GD in unlifted} are run for $T$ iterations. 
\begin{equation}\label{eqn: GD in unlifted}
    \mathbf{x} \leftarrow \mathrm{Proj}_{[-1,1]}(\mathbf{x}+\alpha \mathbf{g}(\mathbf{x}))\:, 
\end{equation}
where $\alpha>0$ is the step size and \textcolor{black}{$\mathrm{Proj}_{[-1,1]}$ is} the projection operator, applied element-wise. Next, we state two properties of the graph Laplacian $\mathbf{L}$.  \par
\begin{lemma}\label{lemm: L is PSD}
For any graph with one connected component, $\mathbf{L}$ is positive semidefinite (PSD).
\end{lemma}
\begin{lemma}\label{lemm: null space of L}
For any graph with one connected component, the null space of $\mathbf{L}$,
\begin{equation}
    \label{eqn: L null space set}
  N(\mathbf{L}) = \{\mathbf{x}\in \mathbb{R}^n : \mathbf{L}\mathbf{x} = \mathbf{0}\},  
\end{equation}
%
is one-dimensional, spanned by the all-ones vector $\mathbf{e}_n$. 
\end{lemma}

Lemma~\ref{lemm: L is PSD} follows from the non-negativity of the objective of \eqref{eqn: cut quco unlifted} as we have $\sum_{(v,u)\in E}(\mathbf{x}_{v}-\mathbf{x}_{u})^2\geq 0$. Lemma~\ref{lemm: null space of L} holds due to the row-sum property of $\mathbf{L}$ (each row or column sums up to $0$) \citep{chung1997spectral}. 

As a direct result of Lemma~\ref{lemm: L is PSD},  $f$ is convex as $\mathbf{L}$ is PSD. A direct result of Lemma~\ref{lemm: null space of L} is that for any $c\in\mathbb{R}$, we have $c\mathbf{e}_n \in N(\mathbf{L})$. 


%
\begin{remark}\label{rem: connetced comp}
    Since we are solving the MaxCut problem, we assume that the graph $G$ has only one connected component, i.e., there is a path from any two nodes. If the graph has multiple connected components, then MaxCut can be solved independently on each component. 
\end{remark}

Although the projection operator in \eqref{eqn: GD in unlifted} does not affect the convexity of $f$, it creates \textbf{fixed-points} that could correspond to different graph cuts. Finding the optimal MaxCut fixed-point(s) is NP-hard in general. 



Due to the absence of a linear term in $f$, any vector in the null space of $\mathbf{L}$ creates a fixed-point that is a stationary point (i.e., $\mathbf{g}(\mathbf{x}) = \mathbf{0}$). Next, we formally define the MaxCut fixed-points of \eqref{eqn: cut quco unlifted}. 
\begin{definition}[MaxCut fixed-points] 
For any graph with one connected component $G=(V,E)$, define the set of MaxCut fixed-points as
%
\begin{align}
    F:=\Big\{ \mathbf{x}\in \{-1,1\}^n \setminus N(\mathbf{L})~~~~~~~ \text{such that} & \nonumber \\ \mathbf{x} = \mathrm{Proj}_{[-1,1]}(\mathbf{x}+\alpha \mathbf{g}(\mathbf{x})) \Big\}\:.
\end{align}\nonumber
That is, $F$ contains binary vectors that are not in the null space of $\mathbf{L}$ but remain unchanged after a gradient ascent step followed by projection. Any point in the interior of the box (i.e., $(-1,1)^n$) \textcolor{black}{cannot be a MaxCut fixed-point because it could be in the null space of $\mathbf{L}$}.
\end{definition}
The above definition indicates that any point in $F$ corresponds to a ``local maximizer'' of \eqref{eqn: cut quco unlifted} in the sense that a projected gradient ascent step returns the exact same point. 




\subsection{\textbf{Lifted} Quadratic Formulation}\label{sec: proposed functions lifted}

Inspired by over-parameterization in machine learning, which is known to improve optimization landscapes and escape \textcolor{black}{\textit{possibly}} suboptimal maxima \citep{du2018power,neyshabur2015path} and since gradient ascent on the unlifted objective \eqref{eqn: cut quco unlifted} often gets stuck at fixed-points and fails to explore other potentially better cuts, we propose a lifted quadratic optimization formulation. 

More specifically, we represent each node $v \in V$ using an $l$-dimensional vector. Here, \textcolor{black}{we term} $l \leq n$ as the lifting parameter. Our proposed Lifted qUadratiC Optimization (LUCO) is: 
\begin{align}\label{eqn: cut luco lifted}
    \max_{\mathbf{X}\in [-1,1]^{n\times l}} \quad h(\mathbf{X}) := \mathrm{tr}(\mathbf{X}^{\top}\mathbf{L}\mathbf{X}) \:, \tag{\texttt{LUCO}}
\end{align}
where the Jacobian is
\begin{equation}
    \label{eqn: jacobian of lifted}
    \mathbf{J}(\mathbf{X}) \in \mathbb{R}^{n\times l} = \mathbf{L}\mathbf{X},
\end{equation}
%
and the update rule of the gradient ascent becomes: 
\begin{equation}\label{eqn: GD in lifted}
    \mathbf{X} \leftarrow \mathrm{Proj}_{[-1,1]}(\mathbf{X}+\alpha \mathbf{J}(\mathbf{X}))\:. 
\end{equation}
Assuming the above updates converge to a fixed-point, the cut value is computed via:
\begin{equation}\label{eqn: cut value for lifted}
\mathbf{z}_v = \mathds{1}\Big\{\sum_{i\in [l]} \mathbf{X}_{v,i}\geq 0\Big\}\:,~~\text{with cut value   } \mathbf{z}^{\top}\mathbf{L}\mathbf{z}\:.
\end{equation}
%
In \eqref{eqn: cut value for lifted}, we check whether the sum of row $v$ in $\mathbf{X}$ is greater than or equal to $0$. 

\begin{remark}
    \textcolor{black}{The gradient updates for each column of $\mathbf{X}$ are equivalent to applying gradient updates in the unlifted formulation. However, the computation of the cut value in \eqref{eqn: cut value for lifted} differs, which leads to different resulting cuts between the unlifted and lifted formulations. Consequently, each column may correspond to a valid solution individually, while their combination yields a different overall solution. This behavior provides the intuition behind the “exploration” effect. Moreover, the lifted formulation naturally enables parallelization, since all columns of the matrix are updated simultaneously, which constitutes the primary motivation for introducing the lifting.  }
\end{remark}


In the following theorem, we characterize the lifted MaxCut fixed-points based on which points do not yield an all-zero Jacobian. The proof is deferred to the Appendix~\ref{sec appen: proof of th 1}. 

%
\begin{theorem}[Lifted MaxCut fixed-points]\label{th: lifted fixed poins}
    Given a graph $G=(V,E)$ with one connected component, its Laplacian matrix $\mathbf{L}$, and the gradient ascent updates in \eqref{eqn: GD in lifted}, then, according to \eqref{eqn: cut value for lifted}, the set of fixed-points that correspond to MaxCut is 
    \begin{align}
     P:= \Big\{ \mathbf{X}\in \{-1,1\}^{n\times l} ~~~~~~\text{~such that} & \nonumber \\ 
      \mathbf{X} \neq \mathbf{e}_n\mathbf{c}^{\top} \land \mathbf{X}\mathbf{e}_n \neq \mathbf{0}_n, \mathbf{c}\in \{-1,1\}^{l\times1} \Big\}\:.
    \end{align}\nonumber
\end{theorem}
This result excludes vectors where all rows of $\mathbf{X}$ are identical (i.e., lie in the null space of $\mathbf{L}$), which would yield a cut value of zero.

\begin{remark}\label{rem: lifted vs. SDP}
Compared to solving the MaxCut SDP in \eqref{eqn: cut SDP}, \eqref{eqn: cut luco lifted} avoids the need to enforce the PSD constraint and the $|V|$ equality constraints. Moreover, the lifting dimension, $l$, does not need to be equal to $n$, as is the case with SDP. In both the MaxCut SDP and our proposed objective in \eqref{eqn: cut luco lifted}, a rounding is needed. Empirically, we will show that QUCO and LUCO are better suited for large-scale graphs where enforcing PSD constraints becomes computationally prohibitive—even when using commercial SDP solvers such as MOSEK \citep{mosek}. 
\end{remark}

\begin{remark}\label{rem: lifted vs. OptGNN}
A recent learning-based method also introduces a lifted version for the MaxCut problem, where the PSD constraints in \eqref{eqn: cut SDP} are replaced by the $\ell_2$ norm on each column of the optimization matrix \citep{opt_gnn}. Then, the authors define a connection between message passing of GNNs \citep{gilmer2017neural} and applying gradient descent on their lifted formulation. There are two differences between our lifted formulation and the one in \citep{opt_gnn}: (i) Our formulation uses only box-constraints, whereas an $\ell_2$ norm is needed in \citep{opt_gnn}; and more importantly (ii) the method in \citep{opt_gnn} embeds trainable weight matrices in every layer that represents the aggregate function of the GNN, and trains over training graphs, whereas our method is not a learning-based approach, avoiding any potential generalizability issues typically encountered in training-data-intensive methods \citep{zhang2017understanding}.
\end{remark}






\section{PROPOSED ALGORITHMS}\label{sec: proposed algs}


In this section, we first introduce two degree-based initialization strategies, followed by the three proposed parallelized algorithms based on the formulations in \eqref{eqn: cut quco unlifted} and \eqref{eqn: cut luco lifted}. Then, we describe the adopted evolution-based hyper-parameter search algorithm. 

\subsection{Degree-based Initialization}\label{sec: algs initial}

We consider two degree-based initialization strategies, where the motivation is that high-degree nodes have greater influence on the cut value when switching between the two partitions.

\subsubsection{Degree-based via Uniform Distribution Initialization (\textbf{DUI}):} 

Here, higher-degree nodes are initialized with values closer to zero than lower-degree nodes. Specifically, we define:
\begin{equation}\label{eqn: direct deg init}
    \mathbf{h}_{\textrm{DUI}} \sim \mathcal{U} \left[ -\left( 1-\frac{\textrm{d}_v}{\Delta} \right), \left(1-\frac{\textrm{d}_v}{\Delta} \right) \right]^n\:,
\end{equation}
    where $\mathcal{U}$ denotes the continuous uniform distribution.

\subsubsection{Importance-based Degree-based Initialization (\textbf{IDI}):} Here, the nodes are first divided into important and unimportant sets based on the graph average degree and $\textrm{d}_v$. Then, two types of node assignments are used to sample from the discrete uniform distribution $\mathcal{U}\{-1,1\}$ to construct $\mathbf{h}_{\textrm{IDI}}$. This approach is inspired by the direct degree-based distribution sampling in \eqref{eqn: direct deg init}, and the local search metaheuristic algorithm in \citep{festa2002randomized,feo1995greedy, feo1989probabilistic}. 

More specifically, we first divide $V$ into an important set  
\begin{equation}
   I := \{v\in V : \textrm{d}_v > \bar{\textrm{d}} + \beta\sigma \}\:, 
\end{equation}
%
and an unimportant set $U = V\setminus I$, where 
\begin{equation}
    \bar{\textrm{d}} = \frac{1}{|V|}\sum_{v\in V} \textrm{d}_v\:, ~ \sigma^2 = \frac{1}{|V|}\sum_{v\in V} (\textrm{d}_v - \bar{\textrm{d}})^2\:,
\end{equation} 
are the mean and variance of the degrees of the graph, respectively. Here, $\beta \in (0,1)$ is a threshold. We initialize the nodes in the important set using the discrete uniform distribution, i.e., for each $i\in I$, $p_i \sim \mathcal{U}\{-1,1\}$. Consequently, we use $p_i$ to initialize the nodes in the unimportant set as follows. First, define indicator vectors $\mathbf{a},\mathbf{b} \in \{0,1\}^n$ such that 
\[ \mathbf{a}_i = \mathds{1}\{p_i=1\}, ~~ \mathbf{b}_i = \mathds{1}\{p_i=-1\} \:.\]


We have $(\mathbf{A}\mathbf{a})_i$ (resp. $(\mathbf{A}\mathbf{b})_i$) indicates the number of the edges from node $i$ to the important (resp. unimportant) nodes. Then, for every node in the unimportant set, i.e., $i\in U$, we have
  \[
    r_i =
    \begin{cases}
      \sim \mathcal{U}\{-1,1\}, 
        & \text{if} ~~~~~(\mathbf{A}\mathbf{a})_i = (\mathbf{A}\mathbf{b})_i,\\
      +1, 
        & \text{if} ~~~~~(\mathbf{A}\mathbf{a})_i < (\mathbf{A}\mathbf{b})_i,\\
      -1, 
        & \text{otherwise}\:,
    \end{cases}
  \]
That is, if the cut gain is tied, choose randomly; otherwise assign $i$ to the side with fewer connections to the important nodes. Based on sets $I$ and $U$, $\mathbf{h}_{\textrm{IDI}}$ is constructed using $p_i$ and $r_i$. 

For the lifted formulation, these initialization strategies are applied to each of the $l$-dimensional column vectors in matrix $\mathbf{X}$. 

\subsection{Parallelized Algorithms}

To improve scalability, we adopt GPU-accelerated batch execution, where $B$ initialization vectors are processed in parallel. This means that multiple vectors from the above initializations are processed at the same time. 

For the first batch, we generate Gaussian-distributed vectors centered at $\mathbf{h}_{\text{DUI}}$ or $\mathbf{h}_{\text{IDI}}$ with covariance matrix $\eta \mathbf{I}$, where $\eta > 0$ is an exploration parameter. For subsequent batches, the mean vector changes to the best binary vector that corresponds to the highest cut value from the previous batch.  

The algorithm for optimizing the unlifted formulation (QUCO) using parallel projected gradient ascent is shown in Algorithm~\ref{alg: QUCO}, referred to as parallelized QUCO or \textbf{pQUCO}. We denote the output of this procedure as $\textrm{pQUCO}(\mathbf{x})$, which returns the best binary vector corresponding to the maximum cut.

Algorithm~\ref{alg: LUCO} presents the procedure of parallelized LUCO (pLUCO). Similar to pQUCO, we denote running Algorithm~\ref{alg: LUCO} by $ \textrm{pLUCO}(\mathbf{X})$. The output of this function is the binary vector with the largest cut from the lifted run.


We note that while the procedures of both Algorithm~\ref{alg: QUCO} and Algorithm~\ref{alg: LUCO} describe the use of projected fixed-step-size gradient ascent (PGA), we use momentum-based PGA in practice based on our empirical results. 

%
\begin{algorithm}[t]
\small
\caption{parallelized QUCO (\textbf{pQUCO})}
\textbf{Input}: Graph $G=(V,E)$, number of initializations $B$ in one batch, number of iterations $T$, step size $\alpha$, and initialization method. Initialize $\mathbf{x}$ and $S_{\textrm{cut}} =\{\cdot\}$ (Empty set to collect MaxCut sets). \\
\vspace{1mm}
\small{1:}  \textbf{For} each of the $B$ vectors, \textbf{Do} (Parallel Execution)  \\
\vspace{1mm} 
\small{2:} \hspace{1mm}  \textbf{For} $T$ iterations, \textbf{Do} (Gradient updates)  \\
\vspace{1mm}
\small{3:} \hspace{6mm}  $\mathbf{x} \leftarrow \mathrm{Proj}_{[-1,1]}(\mathbf{x}+\alpha \mathbf{g}(\mathbf{x}))$\\
\vspace{1mm}
\small{4:} \hspace{1mm} $\mathbf{z}_v = \mathds{1}\{\mathbf{x}_v>0\}$ (Binarization)  \\
\vspace{1mm}
\small{5:} \hspace{1mm} $ S_{\textrm{cut}} \leftarrow S_{\textrm{cut}} \cup \{v\in V : \mathbf{z}_v = 1\} $  \\
\vspace{1mm}
\small{6:} \textbf{return:}  $S^* = \arg \max_{S\in S_{\textrm{cut}}} \textrm{Cut}(S)$   \\
\vspace{1mm}
\vspace{-3.5mm}
\label{alg: QUCO}
\end{algorithm}
%
 
%
\begin{algorithm}[t]
\small
\caption{parallelized LUCO (\textbf{pLUCO})}
\textbf{Input}: Graph $G=(V,E)$, number of initializations $B$ in one batch, number of iterations $T$, step size $\alpha$, lifting parameter $l,$ and initialization method. Initialize $\mathbf{X}$ and $S_{\textrm{cut}} =\{\cdot\}$ (Empty set to collect MaxCut sets). \\
\vspace{1mm}
\small{1:}  \textbf{For} each of the $B$ vectors, \textbf{Do} (Parallel Execution)  \\
\vspace{1mm} 
\small{2:} \hspace{1mm}  \textbf{For} $T$ iterations, \textbf{Do} (Gradient updates)  \\
\vspace{1mm}
\small{3:} \hspace{6mm}  $\mathbf{X} \leftarrow \mathrm{Proj}_{[-1,1]}(\mathbf{X}+\alpha \mathbf{J}(\mathbf{X}))$\\
\vspace{1mm}
\small{4:} \hspace{1mm} $\mathbf{z}_v = \mathds{1}\{\sum_{i\in [l]} \mathbf{X}_{v,i}\geq 0\}$ (Binarization)  \\
\vspace{1mm}
\small{5:} \hspace{1mm} $ S_{\textrm{cut}} \leftarrow S_{\textrm{cut}} \cup \{v\in V : \mathbf{z}_v = 1\} $  \\
\vspace{1mm}
\small{6:} \textbf{return:}  $S^* = \arg \max_{S\in S_{\textrm{cut}}} \textrm{Cut}(S)$   \\
\vspace{1mm}
\vspace{-3.5mm}
\label{alg: LUCO}
\end{algorithm}
%




%
\begin{algorithm}[t]
\small
\caption{parallelized DECO (\textbf{pDECO})}
\textbf{Input}: Graph $G=(V,E)$, number of initializations $B$ in one batch, number of iterations $T$, step size $\alpha$, lifting parameter $l$, and initialization method (DUI or IDI). Initialize $\mathbf{x}$ and $S_{\textrm{cut}} =\{\cdot\}$ (Empty set to collect MaxCut sets). \\
\vspace{1mm}
\small{1:} \textbf{Obtain} $\mathbf{x} \leftarrow \textrm{pQUCO}(\mathbf{x})$ (running pQUCO)\\
\vspace{1mm} 
\small{2:} \textbf{Initialize} columns of $\mathbf{X}$ (DUI or IDI)\\
\vspace{1mm}
\small{3:} $\mathbf{x}\leftarrow \textrm{pLUCO}(\mathbf{X})$ (running pLUCO)\\
\vspace{1mm}
\small{4:} \textbf{Repeat} from Step 1 until the time budget expires\\
\vspace{1mm}
\vspace{-3.5mm}
\label{alg: DECO}
\end{algorithm}
\subsubsection{\textbf{D}imension-Alt\textbf{e}rnating \textbf{Q}uadratic \textbf{O}ptimization Algorithm}

Here, we combine the scalability of the unlifted formulation with the exploratory power of the lifted formulation, we propose an alternating approach: the Dimension Alternating Quadratic Optimization algorithm, or DECO. The intuition is as follows: start with a solution from pQUCO, then lift it to a higher-dimensional representation to refine the cut using pLUCO, and then project back to the unlifted representation. This lift-and-project process can be repeated as long as the run-time budget allows. Algorithm~\ref{alg: DECO} presents the procedure.

\subsection{Evolution-based Hyper-parameters Search}\label{sec: evo-based param search}

Motivated by the need to mitigate manual tuning of parameters, namely the step size $\alpha$ and the number of optimization steps $T$, we adopt an evolution-based parameter search algorithm. The procedure is inspired by the population-based training in Algorithm 1 of \citep{jaderberg2023population}, where combinations of the hyper-parameters are tried, ranked, and then replaced based on function evaluations and a series of evolution rounds. However, we make modifications because in our algorithms, we only search for two parameters ($\alpha$ and $T$). We note that this procedure is applied between batches. 

Based on pre-defined bounds of $\alpha$ and $T$, we randomly select pairs from these to construct a population list that we will try next. 
More specifically, we draw $T$ from the discrete uniform distribution,  $\mathcal{U}\{T_l,T_u\}$. For $\alpha$, we use $10$ to an exponent $e$ that is uniformly sampled from $\mathcal{U}[e_l,e_u]$. 

Based on the number of generations (or evolutions), we first solve for all combinations inside the population, keep the top half and replace the bottom half. The number of evolutions indicate the number of rounds used to update the list of combinations inside the population list. 
Next, we explain the way of replacing the bottom half of combinations that is applied for each round. 

For each combination we want to replace, we perturb a pair from the set of combinations we want to keep and then clip it to the bounded values. More specifically, for $\alpha$, let $e$ be the exponent from a pair that we will keep in the next round. Let $e'$ be the exponent we will include in the next round which is obtained based on a perturbed $e$ as 
\[e' = \textrm{Proj}_{[-4,-1]} [ e + 0.2\epsilon ]\:, ~~\text{where}\quad \epsilon \sim \mathcal{N}(0,1)\:.\]
For the number of iterations, let $T_0$ be some number selected from the combinations we will keep in the next round which will be used to obtain $T'$. Here, $T'$ is the one used to replace some number of steps from the bottom half of the population set. $T'$ is obtained from $T_0$ as
\begin{equation}
  T' = \lfloor T_0 \big( 1+0.2(2\epsilon'-1) \big) \rfloor,  
\end{equation}
%
where $\epsilon' \sim \mathcal{U}(0,1)$, and $\lfloor\cdot\rfloor$ is the integer function.

At the end of the algorithm, the best $\alpha$ and $T$ are selected for the next batch. The impact of adopting this algorithm is given in the Ablation study of Appendix~\ref{sec appen: impact of param search}.

\section{EXPERIMENTAL RESULTS}\label{sec: exp results}
\begin{table*}[t]
\small
\centering
\resizebox{0.99\textwidth}{!}{%
\begin{tabular}{c|cc|cc|cc}
\toprule
\multirow{2}{*}{\textbf{Solver}} & \multicolumn{2}{c}{\textbf{Small ER}} & \multicolumn{2}{c}{\textbf{Gset}} & \multicolumn{2}{c}{\textbf{Large ER}} \\
 & Cut & Time &  Cut& Time &  Cut & Time \\
\midrule
SDP+pGW \citep{williamson2011design} & 23980.67 & 16.67 & -- & -- & -- & -- \\
Gurobi \citep{Gurobi} &  22014.84 & 30.01 & 9628.31 &  82.99 & -- & --\\
CP-SAT \citep{ortools} & 22288.09 & 9.16 & 9630.53 & 53.15 & -- & --\\
\midrule
LwD \citep{ahn2020learning} &  22356.57 &	65.24 & 7876.54 & 167.74 & -- & --\\
ANYCSP \citep{tonshoff2023one} & 21963.42   & 73.55 & 7162.19 & 194.30 & 22759109.40 & 2896.12\\
OptGNN \cite{opt_gnn} &  20385.34 & 109.26 & 6851.25 & 311.17 & -- & -- \\
\midrule
ReSCO \citep{resco} & 22569.92 & 81.65 &  8381.53 & 131.25 &  25220597.48 & 8984.46\\
\midrule
PIGNN \citep{schuetz2022combinatorial} & 22454.29 & 67.74 & 7739.51 & 94.56 & 25223415.89 & 1214.55\\
CRA \citep{pignn_cra} & 22688.29 & 78.43 & 8481.23 & 196.54 &  25230787.14 & 1677.38 \\
\midrule
pQUCO+DUI (Ours) &  20276.83 & 46.88 & 6987.95 & 175.34 & 23032986.33 & 4896.10\\
pQUCO+IDI (Ours)& 22992.16 & 57.35 & 8457.25 & 199.23 & 25312959.41 & 5577.24 \\
\midrule
pLUCO+DUI (Ours)& 18552.57 & 83.29 & 6835.37 & 185.39 & 21896472.81 & 5631.73\\
pLUCO+IDI (Ours)& 22237.51 & 98.21 & 7678.41 & 232.56 & 23360609.38 & 6475.40\\
\midrule
pDECO+DUI (Ours)& 20853.76 & 216.14 & 7123.27 & 273.10 & 23252076.92 & 9483.70\\
pDECO+IDI (Ours)& 22865.41 & 254.92 & 8846.35 & 352.23 & 25407655.15 & 10972.48\\
\bottomrule
\end{tabular}
}
\vspace{-0.2cm}
\caption{{\textcolor{black}{Main results (average cut value and run-time in seconds) using the \textbf{small ER} (with $n\in \{700,800\}$ with $p=0.15$), \textbf{Gset}, and \textbf{large ER} (with $n\in \{20000, 30000, 40000\}$ and $p = 0.1$) datasets. `DUI' and `IDI' correspond to the two initializations strategies described in Section~\ref{sec: algs initial}.``--'' indicates that a method ran out of memory.}}}
\label{tab: main ALL}
\vspace{-0.05cm}
\end{table*}

We code our algorithms using PyTorch. For baselines, we consider: (i) exact solvers, Gurobi \citep{Gurobi} and CP-SAT \citep{ortools}, for the ILP in \eqref{eqn: cut ILP}; (ii) training-data-intensive methods, LwD \citep{ahn2020learning} and ANYCSP \citep{tonshoff2023one} (reinforcement learning) and OptGNN \citep{opt_gnn} (unsupervised learning); (iii) GNN-based training-data-free methods, PIGNN \citep{schuetz2022combinatorial} and CRA \citep{pignn_cra}; and (iv) the recent GPU-based sampling method ReSCO \citep{resco}. 

Furthermore, we consider a parallelized version of the GW algorithm \citep{goemans1995improved} (which we term by SDP+pGW) that uses the MOSEK solver in \citep{mosek} to first obtain the solution of \eqref{eqn: cut SDP}, then solves several initializations in parallel using batching and GPUs. For most of these methods, we use the default code bases. \textcolor{black}{For LwD and OptGNN, we provide details about their implementations in Appendix~\ref{sec appen: baselines implem details}}. All experiments were conducted on an H100 GPU machine. 


For the initialization of our algorithms (DUI and IDI), we scale the initial vectors down by a large factor (e.g., 10,000), as we empirically observed that this leads to more stable optimization with respect to both the number of steps $T$ and the step size $\alpha$. This observation is consistent with findings in neural network initialization literature, where improper scaling can cause exploding or vanishing gradients, negatively impacting convergence. For example, the well-known AlexNet~\citep{krizhevsky2012imagenet} relied on scaling and normalization techniques to stabilize early training. More formal analyses are given in the  Xavier initialization~\citep{glorot2010understanding} and He initialization~\citep{he2015delving}. These results support our observation that scaling down the initialization improves the stability of gradient-based updates in our setting.

For graph datasets, we consider (i) Erdos-Renyi (ER) small dataset consisting of 128 graphs with 700 to 800 nodes and probability of edge creation $p=0.15$ (which indicates that nearly 15\% of possible edges from the complete graph exist) \citep{pCQO}; (ii) 19 graphs from the well-known dataset, Gset\footnote{\tiny{\url{http://web.stanford.edu/~yyye/yyye/Gset/}}}, with number of nodes ranging from 800 to 2000 and number of edges ranging from 5000 to 40,000; (iii) very large ER graph dataset that consists of 15 graphs with 20,000 to 40,000 nodes and $p=0.1$ (\textcolor{black}{with number of edges ranging from 19,999,000 to 44,998,500}). These graphs were generated by the ER NetworkX random graph generator \citep{hagberg2008networkx}. 

For pLUCO and pDECO, we use $l=2$ as our lifting dimension following the study in Section~\ref{sec appen: impact of lifting parameter}. For the IDI initialization, we use $\beta = 0.2$. For the exploration parameter, we use $\eta = 0.8$. For the momentum in PGA, we use $0.9$. For the adopted evolution-based hyper-parameter search, we use $T_l = 3000$, $T_u = 10000$, $e_l = -4$, and $e_u = -1$. The population size is $6$ and the number of evolution rounds is set to $5$. Our code is available online\footnote{\tiny{\url{https://github.com/aMian-9987/GD_maxcut}}}.


\begin{figure*}[t]
    \centering
    \includegraphics[width=\linewidth]{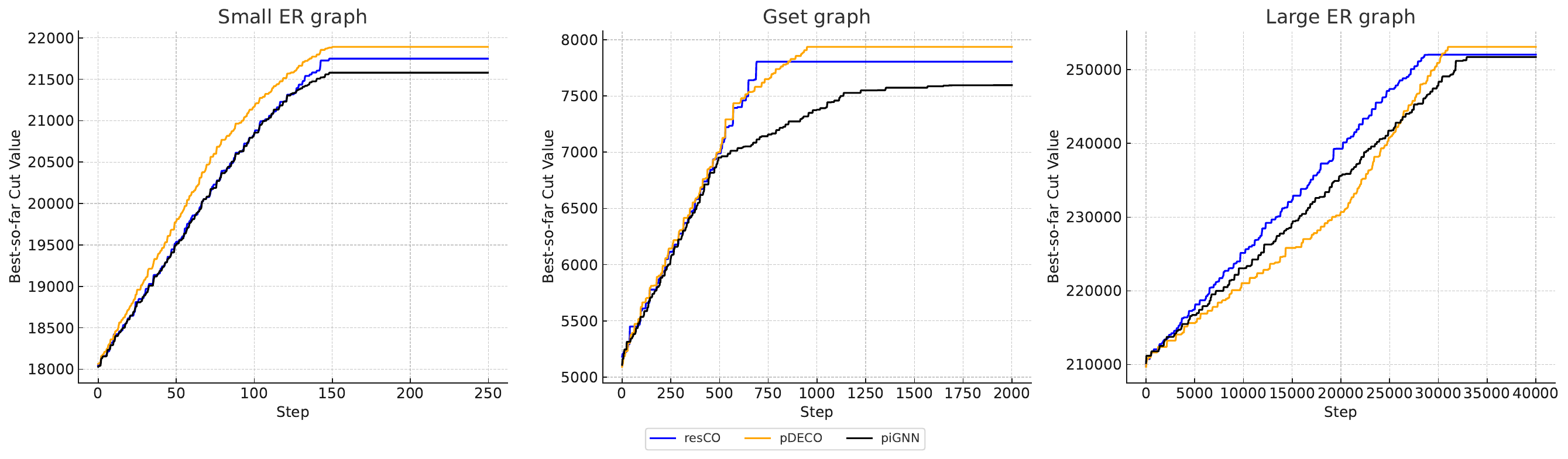}
    \vspace{-.8cm}
    \caption{{Convergence plots of our method (pDECO), ReSCO, and PIGNN. The y-axis corresponds to the cut value of a graph from the small ER dataset (\textit{left}), a graph from the Gset dataset (\textit{middle}), and a graph from a large ER dataset (\textit{right}).} }
    \vspace{-0.0cm}
    \label{fig: conv}
\end{figure*}

\subsection{Main Comparison Results}\label{sec: exp results main results}

Table~\ref{tab: main ALL} present the results for the small ER, Gset, and large ER graph datasets. We note that the entries with ``--'' in the table indicate that this method does not return any results due to running out of memory or requiring a significantly large amount of time given the used compute. 

%
\begin{table*}[t]
\small
\centering
\resizebox{0.99\textwidth}{!}{%
\begin{tabular}{c|cc|cccccc}
\toprule
\multirow{2}{*}{\textbf{Method}} & \multirow{2}{*}{\textbf{Dataset}} & {$(n,p)$}
& \multicolumn{2}{c}{$l=2$} & \multicolumn{2}{c}{$l=3$} & \multicolumn{2}{c}{$l=n$} \\
& & &  Cut Value & Run-time (s) & Cut Value & Run-time (s) & Cut Value & Run-time (s) \\
\midrule
\multirow{4}{*}{\textbf{pLUCO}} &
Small ER & $(700,0.15)$ & 20894.35 & 50.39 & 20951.80 & 57.13 & 22201.47& 68.60 \\
& Small ER & $(800,0.15)$ & 21734.11 & 66.33 & 21805.49 & 71.33 & 22457.81 & 81.45   \\
& large ER & $(20000,0.1)$ & 2330278.41 & 5042.5 & 23347681.55 & 5611.89 & -- & --\\
& large ER & $(30000, 0.1)$ & 24606401.05 & 7077.36 & 24630531.38 & 7735.23& -- & -- \\
\midrule
\multirow{4}{*}{\textbf{pDECO}} &
Small ER & $(700,0.15)$ & 22431.85 & 335.44 & 22793.48  & 354.21 & 22385.71  & 405.41  \\
& Small ER & $(800,0.15)$ & 22589.35 & 347.87 & 23034.17 & 371.64 & 22476.23 & 442.50  \\
& large ER & $(20000,0.1)$ & 25344652.38 & 9560.23 & 25394246.13 & 10581.74 & -- & --  \\
& large ER & $(30000, 0.1)$ &  25731427.59 & 9986.37 & 25814245.25 & 11074.65 & -- & --  \\
\bottomrule
\end{tabular}
}
\vspace{-0.2cm}
\caption{{Ablation study on the lifting parameter, $l\in \{2,3,n\}$, in pLUCO and pDECO. The cut value and run-time results represent the average over the graphs in the datasets given in the second and third columns.}}
\label{tab: ablation lifting}
\end{table*}

In general, all DUI initializations achieve less average cut value than the IDI initialization while requiring less run-time, which is observed on all datasets. This demonstrates the impact of the adopted importance-based initialization. \textcolor{black}{The impact of DUI and IDI versus random initialization is given in Appendix~\ref{sec appen: initialization ablation}}.



\textcolor{black}{For the small ER dataset (i.e., the second and third columns)}, among our methods, pQUCO with IDI initialization achieved an average cut of 22992.16, surpassing all LUCO and DECO variants without search by at least 400. Although SDP+pGW achieved the highest cut value (23980.67) with a runtime of 16.7 s, it does not scale to larger graphs as seen in columns 4 to 7. The reason is that enforcing the PSD constraint in \eqref{eqn: cut SDP} becomes computationally more expensive on larger graphs. Our pDECO+IDI reached 22865.41, only 4.6\% below SDP+pGW. 

Most of our algorithms outperform OptGNN in both cut value and run-time without requiring any training data or a neural network. pDECO+IDI and pQUCO+IDI outperform LwD, also without the need of any training data or an RL agent. 

On the Gset dataset (i.e., the fourth and fifth columns), our pDECO+IDI produced the best cut among all our approaches (8846.35) and improved over pQUCO (which reports 8457.25 at best). This demonstrates the benefit of our dimension-alternating approach in Algorithm~\ref{alg: DECO}. pDECO+IDI also outperformed all of the neural-network-based methods (LwD, OptGNN, PIGNN, and CRA). For Gset, exact solvers (Gurobi and CP-SAT) achieved highest cuts and the lowest run-times. Notably, the ReSCO method based on differentiable sampling achieved 8382 but at a runtime of 131.25 s, whereas pDECO+IDI achieved a much higher cut (884635) at a longer runtime of 352.23 s, indicating that our alternating strategy remains competitive with state-of-the-art GPU-accelerated sampling methods.

For the largest graphs (i.e., the sixth and seventh columns), exact solvers and learning-based methods failed to run (denoted by “–-”). Among GPU-based methods, our pDECO+IDI achieved the highest average cut value, surpassing ReSCO while being more than 2× faster. 

The lifted variants pLUCO+IDI and the unlifted pQUCO+IDI also reached competitive values with significantly less run-time than PIGNN and ReSCO. The dimension-alternating algorithm pDECO consistently offered the best cut size.

\textcolor{black}{In Appendix~\ref{sec appen: additional comparison}, we provide additional results, comparing our algorithms with the Greedy approximation algorithm \citep{sahni1976p} and other baselines using different graph datasets including real-world graph instances from the SNAP Stanford dataset \citep{snapnets}. }

\subsection{Convergence Results}
In Figure~\ref{fig: conv}, we show the convergence of ReSCO, PIGNN, and pDECO (ours) of a graph from each dataset. The results correspond to the best cut value over steps. We selected ReSCO and PIGNN as baselines due to their scalability, similar to our proposed algorithm. 

As observed, for the small ER graph, our method converges faster. In all cases, our pDECO converges to the highest cut value when compared to baselines. \textcolor{black}{Additional convergence plots are given in Appendix~\ref{sec appen: additional convergance}.}

\subsection{Impact of the Lifting Parameter}\label{sec appen: impact of lifting parameter}

Table~\ref{tab: ablation lifting} investigates the effect of the lifting parameter $l$. For small ER graphs, increasing $l$ from 2 to $n$ improved pLUCO cuts but increased the run-time as increasing $l$ means more parameters to optimize. For pDECO, $l=3$ yielded the highest cuts on small ER without excessive runtime, with slight increase for run-time. $l=n$ slightly decreased cut quality. 

On large graphs, results show that very high lifting dimensions were computationally prohibitive, confirming that moderate lifting values achieve the best balance. Therefore, for our main results, we used $l=2$.


\section{CONCLUSION}

We introduced a differentiable lift-and-project framework for the MaxCut problem, built on two continuous relaxations: the unlifted quadratic objective (QUCO) and its lifted version (LUCO). Fixed-points for both formulations were theoretically characterized, with QUCO shown to be convex and to admit MaxCut fixed-points at boundary values, and LUCO characterized by a different set of MaxCut fixed-points. We adopted an importance-based initialization and evolution-based parameter search, combined with a parallel gradient-based optimization strategy and dimension alternation. We evaluated our algorithm on small ER graphs, Gset graphs, real-world graphs, and very large ER graphs, where we demonstrated that the proposed methods achieve competitive cuts values while maintaining practical run-times. For the largest graphs, where exact solvers and learning-based baselines did not run, our parallelized dimension-alternating method achieved the best cut values, outperforming other methods and matching the scalability of very recent SOTA GPU-based sampling methods. Overall, these findings suggest that alternating between unlifted and lifted differentiable formulations, combined with batch parallelization, can enhance exploration without requiring training data, offering a scalable alternative for MaxCut on large graphs.

\bibliographystyle{abbrvnat}
\bibliography{refs.bib}

\section*{Checklist}



\begin{enumerate}

  \item For all models and algorithms presented, check if you include:
  \begin{enumerate}
    \item A clear description of the mathematical setting, assumptions, algorithm, and/or model. [Yes] We provide detailed descriptions in Sections~\ref{sec: proposed functions} and \ref{sec: proposed algs}. 
    \item An analysis of the properties and complexity (time, space, sample size) of any algorithm. [Yes] We provide tun-time comparisons of our algorithms against different baselines. 
    \item (Optional) Anonymized source code, with specification of all dependencies, including external libraries. [Yes] See the footnote in Section~\ref{sec: exp results}.
  \end{enumerate}

  \item For any theoretical claim, check if you include:
  \begin{enumerate}
    \item Statements of the full set of assumptions of all theoretical results. [Yes] 
    \item Complete proofs of all theoretical results. [Yes] We provide one theorem and two lemmas. The full proof of the theorem is given in the first section of the Appendix. The two lemmas are stated from previous work in order to support our claims in Section~\ref{sec: proposed functions relaxed qubo}.
    \item Clear explanations of any assumptions. [Yes]     
  \end{enumerate}

  \item For all figures and tables that present empirical results, check if you include:
  \begin{enumerate}
    \item The code, data, and instructions needed to reproduce the main experimental results (either in the supplemental material or as a URL). [Yes]
    \item All the training details (e.g., data splits, hyperparameters, how they were chosen). [Yes]
    \item A clear definition of the specific measure or statistics and error bars (e.g., with respect to the random seed after running experiments multiple times). [Not Applicable]
    \item A description of the computing infrastructure used. (e.g., type of GPUs, internal cluster, or cloud provider). [Yes]
  \end{enumerate}

  \item If you are using existing assets (e.g., code, data, models) or curating/releasing new assets, check if you include:
  \begin{enumerate}
    \item Citations of the creator If your work uses existing assets. [Yes]
    \item The license information of the assets, if applicable. [Not Applicable]
    \item New assets either in the supplemental material or as a URL, if applicable. [Not Applicable]
    \item Information about consent from data providers/curators. [Not Applicable]
    \item Discussion of sensible content if applicable, e.g., personally identifiable information or offensive content. [Not Applicable]
  \end{enumerate}

  \item If you used crowdsourcing or conducted research with human subjects, check if you include:
  \begin{enumerate}
    \item The full text of instructions given to participants and screenshots. [Not Applicable]
    \item Descriptions of potential participant risks, with links to Institutional Review Board (IRB) approvals if applicable. [Not Applicable]
    \item The estimated hourly wage paid to participants and the total amount spent on participant compensation. [Not Applicable]
  \end{enumerate}

\end{enumerate}















\newpage

\onecolumn
\par\noindent\rule{\textwidth}{1pt}
\begin{center}
{\Large \bf Appendix}
\end{center}
\vspace{-0.1in}
\par\noindent\rule{\textwidth}{1pt}
\appendix



\section{Proof of Theorem~\ref{th: lifted fixed poins}}\label{sec appen: proof of th 1}

\textbf{Re-statement of Theorem~\ref{th: lifted fixed poins}}: Given a graph $G=(V,E)$ with one connected component, its Laplacian matrix $\mathbf{L}$, and the gradient ascent updates in \eqref{eqn: GD in lifted}, then, according to \eqref{eqn: cut value for lifted}, the set of fixed-points that correspond to MaxCut is 
    %
    \begin{align}
     P:= \Big\{ \mathbf{X}\in \{-1,1\}^{n\times l} ~~~\text{~such that} ~~~~ 
      \mathbf{X} \neq \mathbf{e}_n\mathbf{c}^{\top} \land \mathbf{X}\mathbf{e}_n \neq \mathbf{0}_n, \mathbf{c}\in \{-1,1\}^{l\times1} \Big\}\:.
    \end{align}\nonumber
\begin{proof}

We begin the proof by showing that every $\mathbf{X} = \mathbf{e}_n\mathbf{c}^{\top}$ (with $\mathbf{c} \in \mathbb{R}^l$) \textcolor{black}{and every $\mathbf{X}$ that satisfies $\mathbf{X}\mathbf{e}_n = \mathbf{0}_n$} return a cut value of exactly $0$ and therefore is a fixed-point but is not a MaxCut fixed-point. Then, we show that any $\mathbf{X}\in P$ always returns a cut of strictly more than $0$ and its binarized version from \eqref{eqn: cut value for lifted} is always a fixed-point.

We prove the first part as column-by-column. Let \( \mathbf{x} \in [-1,1]^n \) be an arbitrary column of \( \mathbf{X} \), and assume \( \mathbf{L} \mathbf{x} = \mathbf{0} \). Given the objective function of $\eqref{eqn: cut quco unlifted}$, it is clear that the entire sum is zero if and only if each term is zero. That is,
\[
\mathbf{x}_v = \mathbf{x}_u \quad \forall (v,u) \in E.
\]
Because the graph is connected, every node is reachable from every other node via a sequence of edges. Therefore, this condition implies:
\[
\mathbf{x}_1 = \mathbf{x}_2 = \dots = \mathbf{x}_n = c \quad \text{for some } c \in \mathbb{R},
\]
i.e., \( \mathbf{x} = c \mathbf{e}_n \). Applying this to each column \( \mathbf{x}_i, \forall i\in [l] \) of \( \mathbf{X} \), we conclude:
\[
\mathbf{X} = \mathbf{e}_n \mathbf{c}^\top
\quad \text{for some    } \mathbf{c} \in \mathbb{R}^\ell.
\]
Conversely, if \( \mathbf{X} = \mathbf{e}_n \mathbf{c}^\top \), then each column of \( \mathbf{X} \) lies in the null space of \( \mathbf{L} \), and thus:
\[
\mathbf{L} \mathbf{X} = \mathbf{0}_{n\times l}.
\]
Applying the RHS of \eqref{eqn: cut value for lifted} on some $\mathbf{X} = \mathbf{e}_n \mathbf{c}^\top$ results in
\[
\mathbf{z}_v = \sum_{i\in [l]} \mathbf{c}_i\:,
\]
which is independent of $v$ and returns either the vector of all-zero or all-one and hence a cut value of exactly $0$. 

\textcolor{black}{Another case where the cut value is zero is when each row sums up to values that are equal to or more than $1$, which also applies when each row, for all rows, sum up to values that are equal to less than $-1$. }

This characterization of $\mathbf{z}$ is also used for the second part of the proof, where we have a cut value from \eqref{eqn: cut from binary vec} that is only 0 when $\mathbf{z} = \mathbf{e}_n$ or $\mathbf{z} = \mathbf{0}_n$. 
\end{proof}

\section{Impact of the Adopted Parameters Search Algorithm}\label{sec appen: impact of param search}

In this section, we present a study to show the impact of the proposed adoption of the parameters search algorithm discussed in Section~\ref{sec: evo-based param search}. In this study, we compare our pQUCO, pLUCO, and pDECO versus running versions of the algorithms where the parameters (step size $\alpha$ and number of iterations $T$) are manually tuned. We list our selection in Table~\ref{tab: FX values}. Since we run both the unlifted and lifted formulations in pDECO, we include two tuples in the third column. 
%
\begin{table}[htp]
\small
\centering
\resizebox{0.68\textwidth}{!}{%
\begin{tabular}{cccc}
\toprule
\textbf{Dataset} & pQUCO (Algorithm~\ref{alg: QUCO}) & pLUCO (Algorithm~\ref{alg: LUCO}) & pDECO (Algorithm~\ref{alg: DECO}) \\
\midrule
Small ER & $(0.15, 48000)$ & $(0.001, 2500)$ & $(0.10,30000), (0.001, 2000)$ \\
Gset  & $(0.01, 60000)$ & $(0.001, 3000)$ & $(0.012, 80000), (0.001, 2000)$ \\
Large Scale ER & $(0.01, 100000)$ & $(0.001, 5000)$ & $(0.02, 60000), (0.005, 1000)$\\
\bottomrule
\end{tabular}}
\vspace{-0.2cm}
\caption{{The values of $(\alpha,T)$ used for the ``manual'' category of tuning the hyper-parameters in the results of Table~\ref{tab: impact of param search}.}}
\label{tab: FX values}
\end{table}

The results are given in Table~\ref{tab: impact of param search}. As observed, using the adopted evolutionary algorithm for tuning the parameters yields better cut values when compared to fixing the parameters to manually tuned settings. However, we note that, in all datasets, using the search algorithm incurs more run-time. 

\begin{table}[hpt]
\small
\centering
\resizebox{0.98\textwidth}{!}{%
\begin{tabular}{cccccccc}
\toprule
\multirow{2}{*}{\textbf{Method}} & \multirow{2}{*}{\textbf{Selection}} & \multicolumn{2}{c}{Small ER}
& \multicolumn{2}{c}{Gset} 
& \multicolumn{2}{c}{Large ER} \\
  & & Avg. Cut Value & Avg. Run-time (s) & Avg. Cut Value & Avg. Run-time (s) & Avg. Cut Value & Avg. Run-time (s)\\
\midrule
\multirow{2}{*}{pQUCO} & Manual & 22511.44 & 29.26 & 7965.63 & 51.20 & 25019244.21 & 991.52 \\
& Automatic & 22992.16 & 57.35 & 8457.25 & 19.23 & 25312959.41 & 5577.24\\
\midrule

\multirow{2}{*}{pLUCO} & Manual & 22059.23 & 35.40 & 7458.25 & 97.86 & 23284954.22 & 1605.35\\
& Automatic & 22237.51 & 98.21 & 7678.41 & 232.56 & 23360609.38 & 6475.40\\
\midrule

\multirow{2}{*}{pDECO} & Manual & 22181.21 & 91.52 &  8176.00 & 163.45 & 25263415.89 & 4365.99 \\
& Automatic & 22865.41 & 254.92 & 8846.35 & 352.23 & 25407655.15 & 10972.48 \\
\bottomrule
\end{tabular}
}
\vspace{-0.2cm}
\caption{{Average results (Cut values and run-times) of our proposed algorithms using the three datasets with two methods of selecting the hyper-parameters. First is using manually-tuned fixed parameters (denoted as ``Manual''), and the second is using the proposed evolution-based algorithm from Section~\ref{sec: evo-based param search} (denoted as ``Automatic'').}}
\label{tab: impact of param search}
\vspace{-0.00cm}
\end{table}
%


\section{\textcolor{black}{Experimental setting and additional Comparison Results}}\label{sec appen: additional comparison}

In the main experiments (Section~\ref{sec: exp results}), each graph instance is evaluated using a common fixed optimization seed, and the reported results are averaged across different graph instances rather than across repeated runs on the same graph. This protocol evaluates cross-instance performance.
For each selected ER graph, we perform 100 independent runs with different optimization seeds and report the resulting average cut value, thereby isolating variability due to optimization randomness from graph-to-graph variability. Under this setting, we first compare pDECO with the well-known greedy approximation algorithm for MaxCut. We then evaluate pDECO on four large real-world graphs from the SNAP dataset~\citep{snapnets} to further assess its scalability and applicability beyond synthetic graph instances.


\subsection{Comparison with The Greedy MaxCut Algorithm}
\begin{table}[htp]
\small
\centering
\resizebox{0.51\textwidth}{!}{%
\begin{tabular}{cccc}
\toprule
\textbf{Graph} & \textbf{Parameters} & \textbf{pDECO} & \textbf{Greedy Algorithm} \\
\midrule
ER-1 & $(n=700,\; p=0.15)$   & 22793.48    & 22075.94 \\
ER-2 & $(n=800,\; p=0.15)$   & 23034.17    & 22392.52 \\
ER-3 & $(n=20000,\; p=0.10)$ & 10282258.84 & 10043302.65 \\
ER-4 & $(n=30000,\; p=0.10)$ & 23202471.35 & 22861315.29 \\
ER-5 & $(n=40000,\; p=0.10)$ & 40620312.05 & 40398327.80 \\
\bottomrule
\end{tabular}}
\vspace{-0.2cm}
\caption{{
Comparison of the cut values achieved by pDECO (ours) and the MaxCut Greedy algorithm on fixed ER graph instances with varying sizes and edge probabilities. For each instance, results are averaged over 100 independent runs with different optimization seeds.}}
\label{tab: comp with greedy ER}
\end{table}
\begin{table}[htp]
\small
\centering
\resizebox{0.62\textwidth}{!}{%
\begin{tabular}{cccccc}
\toprule
\textbf{Gset Graph} & $(n,m)$ & \textbf{pQUCO} & \textbf{pLUCO} & \textbf{pDECO} & \textbf{Greedy Algorithm} \\
\midrule
G14 & $(800,4694)$     & 3059  & 3052  & 3065  & 3041 \\
G15 & $(800,4661)$     & 3049  & 3044  & 3051  & 3035 \\
G22 & $(2000,19990)$   & 13338 & 13321 & 13361 & 13307 \\
G55 & $(5000,12468)$   & 10288 & 10276 & 10304 & 10169 \\
\bottomrule
\end{tabular}}
\vspace{-0.2cm}
\caption{{Comparison of cut values obtained by our methods (pQUCO, pLUCO, and pDECO) and the greedy algorithm on four benchmark graphs from the Gset dataset.}}
\label{tab: comp with greedy Gset}
\end{table}
Here, we compare our algorithms with the MaxCut greedy algorithm in \cite{sahni1976p}. This algorithm guarantees, in expectation, a cut value of 0.5 from the optimal. The procedure is as follows: Fix an ordering on the vertices in $V$, then, we start with two empty sets $S$ and $\bar{S}$, and put $v_1$ in $S$. For each subsequent vertex, we place it in the set such that $\textrm{Cut}(S)$ is maximized. Other random orderings are used consecutively until the time budget expires. 

Tables~\ref{tab: comp with greedy ER} and ~\ref{tab: comp with greedy Gset} present the results using different ER and Gset graphs. As observed, our algorithms always outperform the greedy algorithm. 

\subsection{Results From the SNAP Real-World Dataset}

In this subsection, we report the results of the pDECO algorithm and baselines using four real-world graph instances from the Stanford Network Analysis Project (SNAP) dataset \citep{snapnets}. 
\begin{table}[htp]
\small
\centering
\resizebox{0.95\textwidth}{!}{%
\begin{tabular}{cccccc}
\toprule
\textbf{Graph} & $(n,m)$ & \textbf{Description} & \textbf{pDECO (Ours)} & \textbf{PIGNN} & \textbf{ResCo} \\
\midrule
email-Eu-core & $(1005,25571)$ & Email data from a large European research institution & 19982 & 19548 & \textbf{20161} \\
ca-GrQc & $(5242,28980)$ & Collaboration network in General Relativity and Quantum Cosmology & 17753 & 17448 & \textbf{17964} \\
musae-facebook & $(22470,171002)$ & Official Facebook pages with mutual likes between sites & \textbf{192874} & 186133 & 175347 \\
ca-HepPh & $(12008,118521)$ & High Energy Physics (Phenomenology) collaboration network & \textbf{80351} & 78732 & 76743 \\
\bottomrule
\end{tabular}}
\vspace{-0.2cm}
\caption{{Comparison of cut values obtained by pDECO, PIGNN, and ResCo on four real-world graphs from the SNAP dataset \citep{snapnets}. The third column provides a description of how each graph was constructed.}}
\label{tab: realworld_pdeco_vs_pignn_resco}
\end{table}
Table~\ref{tab: realworld_pdeco_vs_pignn_resco} presents the results. As observed, pDECO (our algorithm) reports the best cut for the larger graphs (last two rows) as compared to the other training-data-free methods, PIGNN and ReSCO.


\section{Convergence Plots of Two Large ER Graphs w.r.t. Time}\label{sec appen: additional convergance}

\begin{figure*}[htb]
    \centering
    \includegraphics[width=\linewidth]{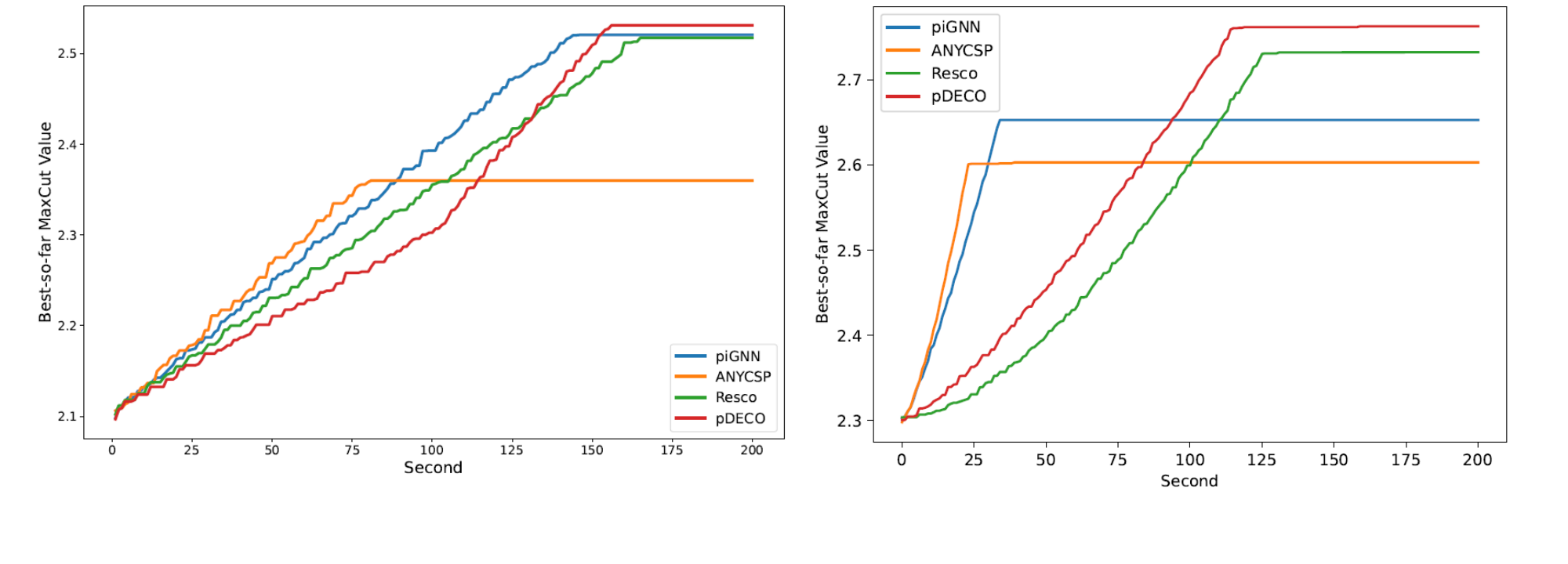}
    \vspace{-1.6cm}
    \caption{{Convergence plots of our method (pDECO), ReSCO, ANYCSP, and PIGNN. The y-axis corresponds to the cut value (multiplied by $10^7$) using two large scale ER graphs.} }
    \vspace{0cm}
    \label{fig: conv 2}
\end{figure*}

\textcolor{black}{In Figure~\ref{fig: conv 2}, we present convergence plots of two large-scale ER graphs w.r.t. execution time (x-axis). Our algorithm, pDECO, is compared against piGNN, ReSCO, and ANYCSP. As observed, pDECO eventually achieves the best cut value. However, ANYCSP and piGNN obtain better results at the early stage of optimization.}


%
\begin{table}[htp]
\small
\centering
\resizebox{0.88\textwidth}{!}{%
\begin{tabular}{ccccc}
\toprule
\textbf{Gset Graph} & $(n,m)$ & pDECO + IDI & pDECO + IDI & pDECO + Random uniform initialization  \\
\midrule
G14 & $(800,4694)$ & 3065	&3059&	3052 \\
G15  & $(800,4661)$ & 3051	&3045	&3038 \\
G22 & $(2000,19990)$ & 	13361&	13343	&13319\\
G55 & $(5000,12468)$ & 10304&	10284&	10227\\
\bottomrule
\end{tabular}}
\vspace{-0.2cm}
\caption{{Cut values using different initialization methods with pDECO on four graphs from the Gset dataset.}}
\label{tab: random init ablation}
\end{table}
\section{Impact of the Proposed/Adopted Initialization}\label{sec appen: initialization ablation}

\textcolor{black}{In this section, we present an ablation study to motive why we needed to use the IDI and DUI initialization strategies. To this end, using our pDECO algorithm (Algorithm~\ref{alg: DECO}), in Table~\ref{tab: random init ablation}, we report a comparison between the proposed DUI and IDI initializations and the random uniform initialization. As observed, our IDI and DUI return the best results.}


\section{Baselines Detailed Implementation}\label{sec appen: baselines implem details}

For every baseline, we use the recommended default setting from their code base other than two methods: LwD and OptGNN. 

For LwD \citep{ahn2020learning}, the code base only include the MIS problem. Therefore, we make the following modifications in the code. Each node's label $s_i \in\{-1,0,+1\}$ represents partition assignment, with 0 denoting undecided. At each stage the dual-head policy outputs (1) a defer decision $d_i \in\{0,1\}$ indicating whether to assign node $i$ now, and (2) an assignment $a_i \in\{+1,-1\}$ if $d_i=1$. The observation per node is $\left[s_i, \mathds{1}\left\{s_i=0\right\}, t / T_{\max }\right]$. Reward is the increment in cut value:
$$
r_t=\operatorname{Cut}\left(s^{(t)}\right)-\operatorname{Cut}\left(s^{(t-1)}\right), \quad \text{where}~~~~ \operatorname{Cut}(s)=\frac{1}{2} \sum_{(i, j) \in E}\left(1-s_i s_j\right) \mathds{1}\left\{s_i \neq 0\right\} \mathds{1}\left(s_j \neq 0\right)
$$
Policy optimization is performed with PPO using the joint log-probability of defer and assignment decisions; the surrogate loss, value loss, and entropy regularization follow standard formulations with clipping. For more information, see the LwD implementation in our code.

For OptGNN \citep{opt_gnn}, since the official implementation does not release any pre-trained weights or model checkpoints, we train the model from scratch following the loss formulation in the original paper. 
We use Adam optimizer with a learning rate of $1\times10^{-3}$ and train for $100$ epochs.

For training data, we construct a large set of synthetic random graphs, since the 
original benchmarks (ER and Gset) are already used for evaluating other baselines 
and should be held out for fair comparison. Specifically:
\begin{itemize}
    \item \textbf{ER graphs}: We generate $1000$ random graphs 
    with node sizes $n \in \{50, 100, 200, 500\}$ and edge probabilities 
    $p \in \{0.1, 0.2, 0.3, 0.5\}$ using the NetworkX implementation. 
    \item \textbf{$d$-regular graphs}: To improve robustness, we additionally 
    generate $200$ $d$-regular graphs with $d \in \{4,8,16\}$. 
\end{itemize}

Validation is performed on a small subset of generated ER and regular graphs 
($5\%$ of the training size). For testing, we strictly hold out 
all the ER and Gset instances that were used by the baselines (127 ER and 80 Gset graphs), 
together with our own $15$ large-scale graphs. 

\end{document}